\documentclass[journal]{IEEEtran}
\ifCLASSINFOpdf
  \usepackage[pdftex]{graphicx}
\else
  \usepackage[dvips]{graphicx}
\fi

\newcommand{\RNum}[1]{\uppercase\expandafter{\romannumeral #1\relax}}

\usepackage{amsfonts}
\usepackage{acronym}
\usepackage{graphicx}
\usepackage{multirow}
\usepackage{cite}
\usepackage{textcomp}
\usepackage[usenames,dvipsnames]{color}
\usepackage[cmex10]{amsmath}
\usepackage[font=footnotesize]{subfig}
\usepackage{amssymb}
\usepackage{balance}
\usepackage{tabularx}
\usepackage{threeparttable}
\usepackage{bbm}
\usepackage{bm}
\usepackage{comment}
\usepackage{url}
\usepackage{fancyhdr}
\usepackage[T1]{fontenc}
\usepackage{type1ec}

\newcommand{\ignore}[1]{}
\begin{document}
\title{Suppressing Alignment: Joint PAPR and Out-of-Band Power Leakage Reduction for OFDM-Based Systems}

%\author{Anas Tom,~\IEEEmembership{Student Member,~IEEE,}
		%Alphan \c{S}ahin,~\IEEEmembership{Member,~IEEE,}
        %H\"{u}seyin Arslan,~\IEEEmembership{Member,~IEEE}        
%\thanks{This work has been}
%of Electrical Engineering, University of South Florida, Tampa,
%FL, 33620 USA e-mail: (atom@mail.usf.edu)}% <-this % stops a space
%\author{\authorblockN{Anas Tom$^1$, Alphan \c{S}ahin$^2$ and H\"{u}seyin Arslan$^{1,3}$\\}
%\authorblockA{$^1$Department of Electrical Engineering, University of South Florida, Tampa, FL, 33620}%\\
%\authorblockA{$^2$Department of Electrical and Computer Engineering, Texas A\&M University, College %Station, TX, 77843}\\
%\authorblockA{$^3$College of Engineering, {\.I}stanbul Medipol University, Beykoz, {\.I}stanbul, 34810
%}\\
%Email: {\tt {atom}@mail.usf.edu}, {\tt alphan@tamu.edu}, {\tt arslan@usf.edu}
%}

\author{\authorblockN{Anas Tom$^1$,~\IEEEmembership{Student Member,~IEEE,} Alphan \c{S}ahin$^1$,~\IEEEmembership{Member,~IEEE}, and H\"{u}seyin Arslan$^{1,2}$,~\IEEEmembership{Fellow,~IEEE}\\}\authorblockA{$^1$Department of Electrical Engineering, University of South Florida, Tampa, FL, 33620}\\
\authorblockA{$^2$College of Engineering, {\.I}stanbul Medipol University, Beykoz, {\.I}stanbul, 34810
}\\
Email: {\tt {atom}@mail.usf.edu}, {\tt alphan@mail.usf.edu}, {\tt arslan@usf.edu}
\thanks{Part of this work has been submitted to ICC 2015 in London.}}

\maketitle
\begin{abstract}
%\boldmath
%\boldmath
%\boldmath
\ac{OFDM} inherently suffers from two major drawbacks: high \ac{OOB} power leakage and high \ac{PAPR}. This paper proposes a novel approach called {\em suppressing alignment} for the joint reduction of the \ac{OOB} power leakage and \ac{PAPR}. The proposed approach exploits the temporal degrees of freedom provided by the \ac{CP}, a necessary redundancy in \ac{OFDM} systems, to generate a suppressing signal, that when added to the \ac{OFDM} symbol, results in marked reduction in both the \ac{OOB} power leakage and \ac{PAPR}. Additionally, and in order to not cause any interference to the information data carried by the \ac{OFDM} symbol, the proposed approach utilizes the wireless channel to perfectly align the suppressing signal with the \ac{CP} duration at the \ac{OFDM} receiver. Essentially, maintaining a \ac{BER} performance similar to legacy \ac{OFDM} without requiring any change in the receiver structure.
\end{abstract}
% \fi
\IEEEpeerreviewmaketitle
\acrodef{OFDM}[OFDM]{Orthogonal frequency division multiplexing}
\acrodef{DFT}[DFT]{discrete Fourier transformation}
\acrodef{IFFT}[IFFT]{inverse fast Fourier transformation}
\acrodef{FBMC}[FBMC]{Filter bank multicarrier}
\acrodef{CP}[CP]{cyclic prefix}
\acrodef{PAPR}[PAPR]{peak-to-average power ratio}
\acrodef{QAM}[QAM]{quadrature amplitude modulation}
\acrodef{OQAM}[OQAM]{offset quadrature amplitude modulation}
\acrodef{FFT}[FFT]{fast Fourier transformation}
\acrodef{RRC}[RRC]{root-raised cosine}
\acrodef{CFO}[CFO]{carrier frequency offset}
\acrodef{SIR}[SIR]{signal-to-interference ratio}
\acrodef{ICI}[ICI]{inter-carrier interference}
\acrodef{ISI}[ISI]{inter-symbol interference}
\acrodef{PPN}[PPN]{ployphase network}
\acrodef{WSSUS}[WSSUS]{wide-sense stationary uncorrelated scattering}
\acrodef{SEM}[SEM]{spectral emission mask}
\acrodef{BWA}[BWA]{broadband wireless access}
\acrodef{BER}[BER]{bit error rate}
\acrodef{CC}[CC]{cancellation carriers}
\acrodef{FCC}[FCC]{Federal Communication Commission}
\acrodef{PSD}[PSD]{power spectral density}
\acrodef{RF}[RF]{radio frequency}
\acrodef{EVM}[EVM]{error vector magnitude}
\acrodef{SNR}[SNR]{signal-to-noise ratio}
\acrodef{SEM}[SEM]{spectral emission mask}
\acrodef{P/S}[P/S]{parallel-to-serial}
\acrodef{S/P}[S/P]{serial-to-parallel}
\acrodef{ACLR}[ACLR]{adjacent channel leakage ratio}
\acrodef{OOB}[OOB]{out-of-band}
\acrodef{SU}[SU]{secondary user}
\acrodef{PU}[PU]{primary user}
\acrodef{Ofcom}[Ofcom]{Office of Communications}
\acrodef{AIC}[AIC]{active interference cancellation}
\acrodef{AWGN}[AWGN]{additive white Gaussian noise}
\acrodef{CSI}[CSI]{channel state information}
\acrodef{LSQI}[LSQI]{least squares with a quadratic constraint}
\acrodef{SOCP}[SOCP]{second-order cone program}
\acrodef{CR}[CR]{cognitive radio}
\acrodef{CCDF}[CCDF]{complimentary cumulative distribution function}
\acrodef{MSE}[MSE]{mean square error}
\acrodef{PDP}[PDP]{power delay profile}
\acrodef{SW}[SW]{subcarrier weighting}
\acrodef{PA}[PA]{power amplifier}
\acrodef{IA}[IA]{interference alignment}
\acrodef{PN}[PN]{pseudo-random}
\acrodef{EAIC}[EAIC]{extended active interference cancellation}
\section{Introduction}
\ignore{\IEEEPARstart{owing}} 
\begin{figure*}
\centering
\includegraphics[width=\textwidth,height=\textheight,keepaspectratio]{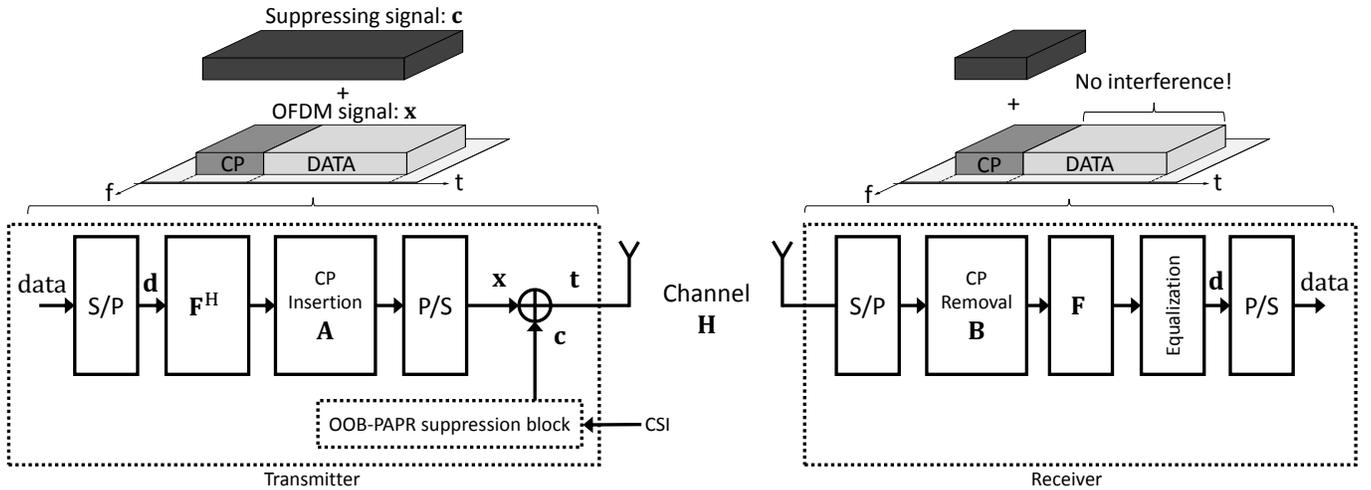}
\caption{System model of an OFDM transmitter and receiver with suppressing alignment.}
\label{Fig:model}
%\vspace{-5mm}
\end{figure*}
\acresetall
\ac{OFDM} is widely regarded as the multicarrier transmission of choice used in most of the existing broadband communication standards, e.g., WiFi, WiMAX, LTE, and IEEE 802.22 WRAN. This prevalent adoption of \ac{OFDM} is due to its numerous advantages such as high spectral efficiency, tolerance to multipath fading, waveform agility, and simple equalization. However, despite all these attractive features, \ac{OFDM} suffers from two major drawbacks: $1)$ \ac{OOB} power leakage as a result of its high spectral sidelobes and $2)$ high \ac{PAPR}. Both of these shortcomings have a large impact on the performance of \ac{OFDM} and can greatly limit its practical applications. For example, the high spectral sidelobes, if not treated, can create severe interference to users operating in adjacent channels \cite{5669234}. The high spectral sidelobes are caused primarily because of the inherent use of rectangular pulse shaping in the generation of \ac{OFDM}, which behaves as a sinc function in the frequency domain with a spectrum that decays as $f^{-2}$ \cite{sahin2013Survey}.
In addition to the high spectral sidelobes, high \ac{PAPR} is another problem that is common to all multicarrier transmission schemes including \ac{OFDM}. The \ac{PAPR} problem arises from the fact that \ac{OFDM} signals are composed of multiple subcarriers with independent amplitudes and phases, that when added together, are more likely to generate a signal with high peak power \cite{4446229}. Such peak power may lead to the signal being severely clipped, especially if it exceeds the linear region of operation of the transmitter \ac{PA}. Signal clipping creates serious inband distortion that ultimately results in large degradation in the \ac{BER} performance at the receiver. Besides the inband distortion, high \ac{PAPR} leads also to spectral spreading, commonly referred to as \ac{OOB} spectral regrowth\cite{749355}.

Both problems received great attention from the research community, where multiple algorithms were proposed that address the two problems either separately such as \cite{Weiss_2004, 6024416, 4450666, 1419026, 1638602, 1557490, 5393050, 1638610, 5353250, 6459499, 5558995, 5672368, 4026715,72_van2009n,6343256} for \ac{OOB} radiation reduction, \cite{1421929, 4446229} and the references therein for \ac{PAPR} reduction, or jointly in \cite{5594133, 4411634, 6735605}. Among the \ac{OOB} reduction techniques, traditional time domain windowing and its variations \cite{Weiss_2004,6024416, 4450666} are considered as a simple and effective way of suppressing the spectral sidelobes, where an extended guard interval is added to the \ac{OFDM} symbol to smooth the transitions between successive symbols. However, despite their simplicity, windowing algorithms reduce the spectral efficiency, especially when the added guard interval is large. Another class of algorithms that are also very useful in reducing the power leakage based on the use of cancellation subcarriers are proposed in \cite{1419026, 1638602, 1557490, 5393050}. There, a subset of subcarriers, modulated with optimized complex weights, are reserved primarily for suppressing the spectral sidelobes of the transmitted signal. However, just like time domain windowing, cancellation algorithms also reduce the transmission rate since the reserved subcarriers are essentially dummy tones that do not carry any data information. More advanced techniques such as precoding are reported in \cite{5353250, 6459499, 5558995, 5672368, 4026715,72_van2009n, 6343256, 1638610}. In general, precoding schemes produce significant reduction in the \ac{OOB} leakage compared to other state of the art algorithms, but they also have their own limitations. For example, the non-orthogonal precoders in \cite{5353250,6459499,72_van2009n, 1638610} destroy the orthogonality between the \ac{OFDM} subcarriers and therefore result in error performance degradation at the receiver. On the other hand, the orthogonal precoders in \cite{5558995, 5672368, 4026715,6343256}, while all are able to maintain the same error performance as plain \ac{OFDM}, they do so by sacrificing the spectral efficiency.

Almost all existing solutions for the \ac{OOB} leakage reduction suffer from either a spectral efficiency loss or \ac{BER} degradation. Furthermore, all of the aforementioned spectral suppression algorithms ignore the issue of high \ac{PAPR}, an inherent characteristic of \ac{OFDM} waveforms. As a result, the gains in \ac{OOB} leakage reduction provided by these algorithms might be misleading, i.e., the spectral sidelobes can potentially grow back up after the high peak power transmitted signal passes through the \ac{PA}. Therefore, and for the reasons outlined above, particularly the spectral regrowth problem, we believe that the best way is to address the two problems jointly as done in \cite{5594133, 4411634, 6735605, 6823751}. Following this path, we herein propose a novel algorithm, that we call {\em suppressing alignment}, for the joint suppression of both the \ac{OOB} leakage and \ac{PAPR}  without any reduction in the transmission rate. Our algorithm exploits the temporal degrees of freedom provided by the \ac{CP}, a necessary redundancy in \ac{OFDM} systems, to properly design a suppressing signal that can effectively reduce both the \ac{OOB} power leakage and \ac{PAPR} of the \ac{OFDM} signal. In particular, our approach adds another dimension to the use of the \ac{CP}. Traditionally, the \ac{CP} has been exploited mainly to mitigate the impact of \ac{ISI} in multipath fading channels. In this work, we extend that functionality by also utilizing the \ac{CP} for the purpose of spectral emissions suppression and \ac{PAPR} reduction. Besides exploiting the \ac{CP}, our design also utilizes the wireless channel to align the generated suppressing signal with the \ac{CP} duration of the \ac{OFDM} symbol at the receiver.  By doing so, the suppressing signal will not cause any interference to the data portion of the \ac{OFDM} symbol. From an interference point of view, the data carried in the \ac{OFDM} symbol appears to be corrupted by the suppressing signal at the transmitter. However, after passing through the channel, the suppressing signal is perfectly aligned with the \ac{CP}. In light of such alignment, the data portion of the \ac{OFDM} symbol appears completely free of interference to the receiver. Thus, after discarding both the \ac{CP} and the aligned suppressing signal through a simple \ac{CP} removal operation, the receiver can decode the data with an error performance similar to that of standard \ac{OFDM}. In addition to maintaining a spectral efficiency and error performance similar to plain \ac{OFDM}, another advantage of our approach is that it does not require any change in the receiver structure of legacy \ac{OFDM}. 

Similar approaches, albeit for different purposes, have previously been proposed in \cite{6451301} for interference alignment in two-tierd networks, in \cite{6516879} for improving the secrecy rate of \ac{OFDM} systems, and very recently in \cite{6914544} for energy harvesting. Nevertheless, we believe that this is the first approach that exploits such a design for the purpose of spectral emissions and \ac{PAPR} containment.

The rest of the paper is organized as follows. In Section \RNum{2}, the system model is introduced. The concept of suppressing alignment and its application to the reduction of \ac{OOB} leakage is presented in Section \RNum{3}. The joint reduction of \ac{OOB} leakage and \ac{PAPR} is presented in Section \RNum{4}. We discuss the practical implementation issues of the proposed approach is Section \RNum{5}. In Section \RNum{6}, we provide the numerical results and finally the conclusion is provided in Section \RNum{7}.

{\em Notations}: $\mathbf{I}_N$ is the $N\times N$ identity matrix; $\mathbf{0}_{N\times M}$ is an all zeros $N\times M$ matrix. The transpose and conjugate transpose are denoted by $\mathbf{(\cdot)}^{\rm T}$ and $\mathbf{(\cdot)}^{\rm H}$, respectively. $\|\cdot \|_2$ denotes the $2$-norm and $\|\cdot \|_\infty$ denotes the uniform norm. $\mathbb{E} [\cdot]$ denotes the expectation operator while $\ker{(\cdot)}$ denotes the kernel of the matrix. The field of real and field of complex numbers are represented by $\mathbbm{R} ~\text{and} ~\mathbbm{C}$, respectively. $\mathcal{CN}(\mu, \Sigma)$ is the complex Gaussian distribution with mean $\mu$ and covariance matrix $\Sigma$.
%\begin{figure*}
%\centering
%%\includegraphics[width=5in,height=2in]{fig3_cropped.pdf}
%\includegraphics[width=\textwidth,height=\textheight,keepaspectratio]{model_v1_cropped.pdf}
%\caption{System model of an OFDM transmitter and receiver with suppressing alignment.}
%\label{Fig:model}
%%\vspace{-5mm}
%\end{figure*}
\section{System Model}
We consider a single link \ac{OFDM} system consisting of a transmitter and a receiver communicating over a Rayleigh multipath channel as shown in \figurename~\ref{Fig:model}. For ease of analysis and without any loss of generality, we assume an adjacent user, employing \ac{OFDM} or any other technology, operating over a bandwidth spanning $K$ subcarriers within the transmission band of the \ac{OFDM} system. Therefore, the \ac{OFDM} transmitter/receiver pair should control their transmissions such that minimal interference is caused to this adjacent user. Let the total number of subcarriers be $N$, where the subcarriers spanning the adjacent user band, i.e., $\{i+1, ..., i+K\},$ are deactivated. The remaining $N_d$ active subcarriers $\{1,..., i\}\cup \{i+K+1,..., N-1\}$, whereas the DC subcarrier is disabled, are modulated by a set of QAM symbols contained in a vector $\mathbf{d} \in \mathbbm{C}^{N_d \times 1}$. To mitigate the effects of \ac{ISI}, a \ac{CP} of length $L$ samples, which is assumed to be larger than the maximum delay spread of the channel, is added to the start of the \ac{OFDM} symbol. The resulting time domain \ac{OFDM} signal is expressed in vectorized form as
\begin{align}
\mathbf{x} = [x_1,...,x_{N+L}]^{\rm T} = \mathbf{AF}^{\rm H}\mathbf{Md}~,
\end{align} 
where $\mathbf{F}$ is the $N$-point \ac{DFT} matrix, $\mathbf{M} \in \mathbbm{R}^{N \times N_d}$ is a subcarrier mapping matrix  containing the $N_d$ columns of $\mathbf{I}_N$ corresponding to the active data subcarriers and $\mathbf{A} \in \mathbbm{R}^{(N+L)\times N}$ is the \ac{CP} insertion matrix defined as
\begin{align*}
\mathbf{A=}
\begin{bmatrix}
\mathbf{0}_{L\times N-L} & \mathbf{I}_L \\ ~~~~~\mathbf{I}_N
\end{bmatrix}~.
\end{align*}
To control the spectral emissions of the transmitted signal as well as its \ac{PAPR}, the \ac{OOB}-\ac{PAPR} suppression block generates a time-domain {\em suppressing signal} $\mathbf{c} = [c_1,...,c_{N+L}]^{\rm T}$ with the same length as the \ac{OFDM} signal, i.e., $\mathbf{c \in \mathbbm{C}}^{(N+L)\times 1}$. Furthermore, let the suppressing signal $\mathbf{c}$ be expressed as
\begin{align}
\mathbf{c=Ps}~,
\end{align}
where $\mathbf{P}\in \mathbbm{C}^{(N+L)\times L}$ and $\mathbf{s}\in \mathbbm{C}^{L\times 1}$. The transmitted signal is then given as
\begin{align}
\mathbf{t=x+c=AF}^{\rm H}\mathbf{Md+Ps}~.
\label{Eq:tx_signal}
\end{align}
The design of both $\mathbf{s}$ and $\mathbf{P}$ will be discussed in detail in the following section; however for the time being, it suffices to say that $\mathbf{c=Ps}$ will be designed to suppress both the spectral sidelobes and \ac{PAPR} of the transmitted signal.
\section{Suppressing Alignment}
In this section, we introduce the concept of suppressing alignment and discuss its use in suppressing the spectral emissions of the transmitted \ac{OFDM} signal. The application of suppressing alignment in reducing the \ac{PAPR} will be discussed in the next section. Our main aim in this section is to construct the suppressing signal $\mathbf{c}$ in \eqref{Eq:tx_signal} so that the transmitted signal has better spectral emissions compared to conventional \ac{OFDM} signals. More specifically, the suppressing signal $\mathbf{c}$ or equivalently $(\mathbf{Ps})$ is designed under two goals in mind: 1) to minimize the \ac{OOB} power leakage of the transmitted signal in the adjacent band and 2) to avoid causing any interference to the information data carried by the \ac{OFDM} symbol, in the sense that the receiver is able to recover all information data sent by the transmitter. In the subsequent discussion, the vector $\mathbf{s}$ will be designed to fulfill the first requirement while the matrix $\mathbf{P}$ is designed to satisfy the latter. 

We first consider the construction of the matrix $\mathbf{P}$. Since the suppressing signal is added to the \ac{OFDM} signal before transmission in \eqref{Eq:tx_signal}, the information data carried by the \ac{OFDM} signal is distorted and the receiver might not be able to recover the data if the suppressing signal is not properly designed. To achieve such proper design, we need to examine the received signal at the receiver after passing through the channel. 

Let the channel between the transmitter and receiver be an i.i.d. Rayleigh fading channel represented by the vector $\mathbf{h=}[h_0,...,h_l] \sim \mathcal{CN}(0, \mathbf{I}_{l+1}/(l+1))$. We can then express the received signal as
\begin{align}
\mathbf{r=Ht+n}~,
\label{Eq:received_signal}
\end{align}  
where $\mathbf{H}\in \mathbbm{C}^{(N+L)(N+L)}$ is a Toeplitz matrix used to model the convolution between the transmitted signal $\mathbf{t}$ and the channel $\mathbf{h}$ and is given by
\begin{align}
\mathbf{H}=
\begin{bmatrix}
h_0 & 0 & \cdots & 0 & h_l & \cdots & h_1\\
\vdots & \ddots & \ddots & \ddots & \ddots & \ddots & \vdots\\
\vdots & \ddots & \ddots & \ddots & \ddots & \ddots & h_l\\
h_l & \cdots & \cdots & h_0 & 0 & \cdots & 0\\
0 & \ddots & \ddots & \ddots & \ddots & \ddots & \vdots\\
0 & \ddots & 0 & h_l & \cdots & \cdots & h_0 
\end{bmatrix}~,
\label{Eq:chann_toeplitz}
\end{align}
and $\mathbf{n} \in \mathbbm{C}^{(N+L)\times 1} \sim \mathcal{CN}(0,\sigma^2 \mathbf{I}_{N+L})$ is an \ac{AWGN} vector. Assuming perfect synchronization and after the \ac{S/P} conversion, the receiver removes the first $L$ \ac{CP} samples and then applies \ac{DFT}. Using \eqref{Eq:tx_signal}, the received signal after \ac{CP} removal and \ac{DFT} operation can be written as
\begin{align}
\mathbf{y=FBHt+\bar{n}=FBHAF}^{\rm H}\mathbf{Md+FBHPs+\bar{n}}~,
\label{Eq:received_signal_after_cp_removal}
\end{align}
where $\mathbf{B} \in \mathbbm{R}^{N\times (N+L)}$ is the \ac{CP} removal matrix and $\mathbf{\bar{n}} \in \mathbbm{C}^{N\times 1}$ is a noise vector obtained after removing the first $L$ samples from $\mathbf{n}$ and applying the \ac{DFT}. 

We are now ready to address the design of the matrix $\mathbf{P}$ by examining \eqref{Eq:received_signal_after_cp_removal}. As stated before, our goal in designing $\mathbf{P}$ is that the interference caused by the added suppressing signal should be zero at the receiver. Therefore, the following must hold true
\begin{align}
\mathbf{FBHPs=0}~.
\label{Eq:null}
\end{align}
If \eqref{Eq:null} is satisfied, then, the received vector $\mathbf{y}$ in \eqref{Eq:received_signal_after_cp_removal} becomes similar to legacy \ac{OFDM} received data and the receiver would be able to apply single-tap equalization to recover the information symbols. Essentially, the information data in the vector $\mathbf{d}$ experiences zero interference from the suppressing signal.

 Assuming perfect \ac{CSI} at the transmitter, it is clear from \eqref{Eq:null} that if $\mathbf{P}$ belongs to the null-space of the matrix $\mathbf{BH}$, i.e., $\ker{\mathbf{(BH)}}$, then \eqref{Eq:null} is satisfied regardless of the value of the vector $\mathbf{s}$. Using the rank-nullity theorem\ignore{\cite{Strang:1976}}, the dimension of the null-space of $\mathbf{BH}\in \mathbbm{C}^{N\times (N+L)}$ is obtained as $\dim{(\ker{\mathbf{(BH)}})} = N+L - \text{rank}(\mathbf{(BH)}) = L$, since $\text{rank}(\mathbf{BH})= N$. Hence, by choosing $\mathbf{P}$ such that its columns span $\ker{\mathbf{(BH)}}$, the condition in \eqref{Eq:null} is satisfied and the receiver can recover the data using legacy \ac{OFDM} reception. Accordingly, we design $\mathbf{P}$ such that
\begin{align}
\text{span}\mathbf{(P)=\ker{(BH)}}~,
\label{Eq:span_BH}
\end{align}
which is accomplished by choosing the columns of $\mathbf{P}$ as an orthogonal basis of $\ker{\mathbf{(BH)}}$. Using the singular value decomposition, $\mathbf{BH}$ can be factored as
\begin{align}
\mathbf{BH=U\Sigma V}^{\rm H}~,
\end{align}
where $\mathbf{U} \in \mathbbm{C}^{N\times N}$, $\mathbf{\Sigma} \in \mathbbm{C}^{N\times (N+L)}$ is a diagonal matrix holding the singular values of $\mathbf{BH}$, and $\mathbf{V} \in \mathbbm{C}^{(N+L)\times (N+L)}$. If $\mathbf{V}$ is expressed as
\begin{align*}
\mathbf{V=}[\mathbf{v}_0~\mathbf{v}_1~...~\mathbf{v}_{N+L-1}]~,
\end{align*}
then the last $L$ columns of $\mathbf{V}$ constitute an orthogonal basis that spans the null-space of $\mathbf{(BH)}$. Therefore, $\mathbf{P}$ is chosen as
\begin{align}
\mathbf{P=}[\mathbf{v}_{N}~\mathbf{v}_{N+1}~...~\mathbf{v}_{N+L-1}]~.
\label{Eq:p_design}
\end{align}
We remark that such construction of $\mathbf{P}$ allows interference-free transmission and is in principle similar to \ac{IA} \cite{4567443}. In particular, $\mathbf{P}$ aligns the interference from the suppressing signal to the portion of the \ac{OFDM} symbol spanned by the \ac{CP} as shown in \figurename~\ref{Fig:model}. 

We now consider the design of the vector $\mathbf{s}$. Before we go into the details of our proposed method, let's first examine the interference caused by the transmitted signal \eqref{Eq:tx_signal} over the $K$ subcarriers occupied by the user in the adjacent band. The signal spectrum of the transmitted signal \eqref{Eq:tx_signal} is given as
\begin{align}
\bm{\mathcal{S}}_t=\mathbf{F}_{\zeta N, \beta}\mathbf{(AF}^{\rm H}\mathbf{Md+Ps)}~,
\label{Eq:total_spectrum}
\end{align}
where $\zeta$ is the upsampling factor, i.e., $\zeta$ samples per subcarrier are considered, $\beta=N+L$, and $\mathbf{F}_{\zeta N, \beta}$ is an $\zeta N \times \beta$ \ac{DFT} matrix. Using \eqref{Eq:total_spectrum}, the interference in the adjacent band can be given as
\begin{align}
\bm{\mathcal{I}}_K=\bm{\mathcal{F}}_K\mathbf{(AF}^{\rm H}\mathbf{Md+Ps)} = \underbrace{\bm{\mathcal{F}}_K\mathbf{AF}^{\rm H}\mathbf{Md}}_{\bm{\mathcal{F}}_\mathbf{d}} + \underbrace{\bm{\mathcal{F}}_K\mathbf{P}}_{\bm{\mathcal{F}}_\mathbf{s}}\mathbf{s}~,
\label{Eq:interference}
\end{align}
where $\bm{\mathcal{F}}_K$ is a sub-matrix of $\mathbf{F}_{\zeta N, \beta}$ containing only the rows that correspond to the subcarriers occupied by the adjacent user. The first term in \eqref{Eq:interference} represents the \ac{OOB} power leakage from the information data and the second term is the \ac{OOB} power leakage from the suppressing signal $\mathbf{c}$. To minimize the interference power in the adjacent band, we calculate $\mathbf{s}$ such that
%\begin{equation}
%\begin{aligned}
%& \mathbf{s=\text{arg}\min_s{\|\bm{\mathcal{F}_d+\mathcal{F}_ss}\|_2}}, \\
%& \text{subject to}
%~~\mathbf{\|s\|_2^2 \leq \epsilon},
%\label{Eq:optimization}
%\end{aligned}
%\end{equation}
\begin{equation}
\begin{aligned}
& \mathbf{s=\text{arg}\min_s}\mathbf{\|\bm{\mathcal{F}}_d+\bm{\mathcal{F}}_ss\|}_2~~\text{subject to}~~ \mathbf{\|s\|}_2^2 \leq \epsilon~,
%& \text{subject to}
%~~\mathbf{\|s\|}_2^2 \leq \epsilon~,
\label{Eq:optimization}
\end{aligned}
\end{equation}
where $\epsilon$ is a power constraint on the vector $\mathbf{s}$ to avoid spending too much power on the suppressing signal. We note here that the power of the suppressing signal $\mathbf{c}$ is equal to the power of the vector $\mathbf{s}$ since $\mathbf{P}$ is an orthogonal matrix. The optimization problem in \eqref{Eq:optimization} is known as a \ac{LSQI} problem. To solve this problem, we first consider the unconstrained least squares problem, i.e., without the power constraint. The solution to the least squares problem is
\begin{align}
\mathbf{s = -(\bm{\mathcal{F}_s}}{\rm ^H}\mathbf{\bm{\mathcal{F}_s}})^{-1}\bm{\mathcal{F}_s}{\rm ^H} \mathbf{\bm{\mathcal{F}_d}}~.
\label{Eq:least_squares}
\end{align}
It is clear that the calculated $\mathbf{s}$ in \eqref{Eq:least_squares} is also the solution to the problem in \eqref{Eq:optimization} if $\mathbf{\|s\|}_2^2\leq \epsilon$, and in this case we have an analytical solution. However, if $\mathbf{\|s\|}_2^2\geq \epsilon$, then there is no analytical solution and in order to solve the problem, we have to consider the following unconstrained problem
\begin{align}
\mathbf{s=\text{arg}\min_s}{\|\mathbf{\bm{\mathcal{F}_d}+\bm{\mathcal{F}_ss}}\|_2}+\lambda_0\mathbf{\|s\|}_2^2~,
\end{align}
where $\lambda_0 > 0$ is the Lagrange multiplier. The solution in this case is given by
\begin{align}
\mathbf{s} = -(\mathbf{\bm{\mathcal{F}_s}}{\rm ^H}\mathbf{\bm{\mathcal{F}_s}}+\lambda_0 \mathbf{I})^{-1}\mathbf{\bm{\mathcal{F}_s}}{\rm ^H}\mathbf{\bm{\mathcal{F}_d}}~.
\label{Eq:exp_s}
\end{align}
For a proper Lagrange multiplier, which can be found using the bi-section search algorithm \cite{Boyd:2004:CO:993483}, $\mathbf{\|s\|}_2^2=\epsilon$.
Alternatively, \eqref{Eq:optimization} can be solved numerically using any of the publicly available optimization solvers. 
\section{Joint \ac{PAPR} and \ac{OOB} Power Leakage Reduction}
\ac{PAPR} is an important metric for multi-carrier systems. Any increase in the \ac{PAPR} might drive the power amplifier at the transmitter to operate in the non-linear region. This can potentially cause spectral regrowth in the sidelobes, erasing any \ac{OOB} reduction gains achieved before the power amplifier. Therefore, as an extension to the results in the previous section, we propose to jointly minimize the \ac{PAPR} and \ac{OOB} power leakage to avoid such problem. 

The \ac{PAPR} of the transmitted signal is the ratio of the maximum instantaneous power to the average power which is given as
%\begin{align}
%\text{\ac{PAPR}}=\frac{\underset{0 \leq n \leq \rho(N+L-1)}{\max{[|t(n)|%^2]}}}{\mathbb{E}[|t(n)|^2]}=\frac{\mathbf{\|x+Ps\|_\infty}}{\frac{1}%{\rho(N+L)}\mathbf{\|x+Ps\|_2^2}}
%\end{align}
\begin{align}
\text{\ac{PAPR}} = \frac{\mathbf{\|t\|}^2_\infty}{\frac{1}%
{(N+L)}\mathbf{\|t\|}_2^2}=\frac{\mathbf{\|x+Ps\|}^2_\infty}{\frac{1}%
{(N+L)}\mathbf{\|x+Ps\|}_2^2}~.
\end{align}
Accordingly, to minimize the \ac{OOB} interference as well as the \ac{PAPR}, we extend the optimization problem in \eqref{Eq:optimization} as follows
\begin{equation}
\begin{aligned}
& \mathbf{s=\text{arg}\min_s}{(1-\lambda)\mathbf{\|\bm{\mathcal{F}_d+\mathcal{F}_ss}\|}_2 + \lambda\mathbf{\|x+Ps\|}_\infty}~, \\
& \text{subject to}
~~\mathbf{\|s\|}_2^2 \leq \epsilon~,
\label{Eq:joint}
\end{aligned}
\end{equation}
where the weighting factor, $\lambda \in [0,1]$, is for controlling the amount of minimization for both \ac{OOB} power leakage and \ac{PAPR}. This adaptation parameter can be adjusted to emphasize one problem over the other depending on the system design requirements. For example, when $\lambda = 0$, the objective function turns into a pure \ac{OOB} power leakage reduction problem and \eqref{Eq:joint} is equivalent to \eqref{Eq:optimization}. On the other hand, \eqref{Eq:joint} is a pure \ac{PAPR} reduction problem when $\lambda = 1$. Similar to \eqref{Eq:optimization}, the amount of power consumed by the suppressing signal is controlled by $\epsilon$. 

Both the objective function and the constraint in \eqref{Eq:joint} are convex which renders the problem as a convex optimization problem that can be solved numerically by any convex optimization solver. In this work, we utilize YALMIP \cite{1393890}, a free optimization package that is easily integrated with MATLAB, and MOSEK \cite{msk} as the underlying solver to obtain a numerical solution to \eqref{Eq:joint}.
\section{Practical Implementation Issues}
\subsection{Imperfect channel estimation}
In practice, the assumption of perfect channel knowledge at the transmitter might not be valid. In this subsection, we analyze the performance of the proposed algorithm when the transmitter has imperfect \ac{CSI}. The channel is estimated at the receiver and the \ac{CSI} is fed back to the transmitter. The transmitter then uses this CSI to generate the suppressing signal $\mathbf{c=Ps}$. To evaluate the impact of channel estimation errors, we assume that the channel known at the transmitter is different than the actual channel that the signal is transmitted through. We model the noisy channel estimation as
\begin{align}
\hat{\mathbf{H}} = \mathbf{H+}\mathbf{E}~,
\label{Eq:error_chan}
\end{align}
where $\mathbf{E} = \sigma_e \mathbf{\Omega}$ is the channel error matrix and $\mathbf{\Omega}$ is Toeplitz with the same structure as \eqref{Eq:chann_toeplitz}. The non-zero entries of $\mathbf{\Omega}$ are i.i.d. complex Gaussian with zero mean and unit variance. We quantify the error in channel estimation by the \ac{MSE} $\sigma_e^2$ defined as \cite{4133864}
\begin{align}
\sigma_e^2 = \frac{\mathbb{E}[|\hat{h_{ij}}-h_{ij}|^2]}{\mathbb{E}[|h_{ij}|^2]}~.
\end{align}

The received signal after \ac{CP} removal and \ac{DFT} operation is given by \eqref{Eq:received_signal_after_cp_removal}, where the precoding matrix $\mathbf{P}$ is designed based on knowledge of the channel at the transmitter. If the channel $\mathbf{H}$ communicated back to the transmitter by the receiver is erroneous, then $\mathbf{P}$ is designed based on $\mathbf{H}$ as opposed to the true channel $\mathbf{\hat{H}}$. Therefore, the second term in \eqref{Eq:received_signal_after_cp_removal} would not vanish, i.e., \eqref{Eq:null} is not true any more. This effectively means that the suppressing signal leaks into the data part of the \ac{OFDM} symbol instead of precisely being aligned with the \ac{CP} duration. Nevertheless, the erroneous channel information does not effect the \ac{OOB} power leakage and \ac{PAPR} reduction performance of the proposed method, since the suppressing signal is still designed based on \eqref{Eq:optimization} or \eqref{Eq:joint}.

 The average power leakage of the suppressing signal into the data part of the received \ac{OFDM} symbol can be expressed as
 \begin{equation}
\begin{aligned}
\xi = \frac{1}{N}\mathbb{E}[\mathbf{\|B\hat{H}Ps\|}^2_2]~.
\label{Eq:leaked_pow}
\end{aligned}
\end{equation}
To evaluate the above the expression, we utilize the closed-form expression for $\mathbf{s}$ in \eqref{Eq:exp_s}, which after substituting $\mathbf{\bm{\mathcal{F}_d}}$ from \eqref{Eq:interference}, can be written as
\begin{align}
\mathbf{s} = \underbrace{-(\mathbf{\bm{\mathcal{F}_s}}{\rm ^H}\mathbf{\bm{\mathcal{F}_s}}+\lambda_0 \mathbf{I})^{-1}\mathbf{\bm{\mathcal{F}_s}}{\rm ^H}\bm{\mathcal{F}}_K\mathbf{AF}^{\rm H}\mathbf{M}}_{\mathbf{\Phi}}\mathbf{d} = \mathbf{\Phi d}~.
\end{align}
Substituting $\hat{\mathbf{H}}$ from \eqref{Eq:error_chan} and the above expression for $\mathbf{s}$, the mean leaked power in \eqref{Eq:leaked_pow} can now be evaluated as 
\begin{align*}
\xi &= \frac{1}{N}\mathbb{E} \big[\mathrm{tr}[(\mathbf{B(H+E)P\Phi d})^{\rm H}(\mathbf{B(H+E)P\Phi d})]\big]\\
&= \frac{1}{N}\mathbb{E}\big[ \mathrm{tr}[\mathbf{d}^{\rm H}\mathbf{\Phi}^{\rm H}\mathbf{P}^{\rm H}(\mathbf{H+E})^{\rm H}\mathbf{B}^{\rm H}\mathbf{B(H+E)P\Phi d}] \big]\\
&= \frac{1}{N}\mathrm{tr}\big[\mathbf{\Phi}^{\rm H}\mathbf{P}^{\rm H}\mathbb{E}[(\mathbf{H+E})^{\rm H}\mathbf{B}^{\rm H}\mathbf{B(H+E)]P\Phi}\mathbb{E}[\mathbf{d}\mathbf{d}^{\rm H}]\big ]
\end{align*}
Since $\mathbf{BHP=0}$ and the data vector $\mathbf{d}$ is assumed to have zero mean and covariance $\mathbb{E}[\mathbf{d}\mathbf{d}^{\rm H}] = \mathbf{I}_{N_d}$, we arrive at
\begin{align}
\xi &= \frac{1}{N}\mathrm{tr}\big[\mathbf{\Phi}^{\rm H}\mathbf{P}^{\rm H}\mathbb{E}[\mathbf{E}^{\rm H}\mathbf{B}^{\rm H}\mathbf{BE]P\Phi}\big ] \\&= \frac{1}{N}\mathrm{tr}\big[ \mathbb{E}[\mathbf{EP\Phi}\mathbf{\Phi}^{\rm H}\mathbf{P}^{\rm H}\mathbf{E}^{\rm H}]\mathbf{B}^{\rm H}\mathbf{B} \big]~.
\end{align}
Let $\mathbf{Z}=\mathbf{P\Phi}\mathbf{\Phi}^{\rm H}\mathbf{P}^{\rm H}$, $\mathbf{Y=EZE}^{\rm H}$ and the projection matrix $\mathbf{G}=\mathbf{B}^{\rm H}\mathbf{B}$ defined as
\begin{align*}
\mathbf{G=}\begin{bmatrix}
\mathbf{0}_{L\times L} & \mathbf{0}_{L\times N}\\ \mathbf{0}_{N\times L} & \mathbf{I}_{N\times N}
\end{bmatrix}~.
\end{align*}
Thus,
\begin{align}
\xi=\frac{1}{N}\mathrm{tr}\big[ \mathbb{E}[\mathbf{Y}] \mathbf{G} \big]~.
\label{Eq:leaked_simp}
\end{align}
Using the definition for $\mathbf{Y}$ above as well as the Toeplitz property of the error matrix i.e., $E_{ij}=e_{i-j}$, the expectation in \eqref{Eq:leaked_simp} can now be evaluated as
\begin{align}
\mathbb{E}[Y]_{ij} = \sum_{kl}\mathbb{E}[E_{ik}Z_{kl}E^*_{jl}] = \sum_{kl}\mathbb{E}[e_{i-k}Z_{kl}e^*_{j-l}]~,
\end{align}
and since $\mathbb{E}[e_ie^*_j]=\frac{1}{L}\sigma_e^2\delta'_{ij}$, we then have
\begin{align}
	\mathbb{E}[Y_{ij}] = \frac{1}{L}\sigma_e^2\sum_{kl}Z_{kl}\delta'_{i-k,j-l}~.
\end{align}
Due to the structure of the projection matrix $\mathbf{G}$, it only selects entries with $i=j=L+1, L+2, ..., L+N$. Accordingly, we arrive at the final expression for the leaked power as
\begin{align}
\xi&=\frac{1}{N}\mathrm{tr}\big[ \mathbb{E}[\mathbf{Y}] \mathbf{G} \big]=\frac{1}{LN}\sigma_e^2\sum_{i=L+1}^{N+L}\sum_{k,l=1}^{N+L}Z_{kl}\delta'_{i-k,i-l}\\
&= \frac{1}{LN}\sigma_e^2\sum_{k=1}^{N+L}Z_{kk}\Psi_k ~,
\label{Eq:leaked_closed_form}
\end{align}
where $\Psi_k=\sum_{i=L+1}^{N+L}\delta'_{i-k,i-k}$ and is equal to
\begin{align}
\Psi_k = \begin{cases}
k-1~, & 1 \leq k \leq L~\\
L~, & L+1 \leq k \leq N+1~\\
N+L-k+1~, & N+2 \leq k \leq N+M~\\
0~, & \text{otherwise}
\end{cases}~.
\end{align}
It is worth noting that the closed form expression in \eqref{Eq:leaked_closed_form} represents the power leakage when we consider only the \ac{OOB} reduction but not the joint reduction of PAPR and OOB since a closed-form solution for the suppressing signal does not exist in the latter case. Alternatively, the power leakage in the case of joint reduction of \ac{OOB} and PAPR is evaluated through simulation.
\subsection{Synchronization}
Another important factor for proper operation of the proposed method is time synchronization. It is very critical to know the start of the transmitted frame in order to guarantee exact alignment of the suppressing signal and zero interference to the information symbols. Synchronization in \ac{OFDM} systems is usually achieved by either transmitting a known training sequence (preamble) \cite{650240} or by exploiting the redundancy of the \ac{CP} \cite{4595664, 497156}. Preamble based synchronization algorithms can be incorporated easily with our proposed approach, where the suppressing signal is not generated during the synchronization phase. However, this absence of the suppressing signal during the synchronization phase will not have any detrimental effects on the \ac{OOB} interference or \ac{PAPR} since the preamble is usually made up of \ac{PN} sequences that have low \ac{OOB} leakage and \ac{PAPR} \cite{796380}.

\ac{CP}-based synchronization is based on the fact that the \ac{CP} samples are similar to the corresponding data samples at the end of the \ac{OFDM} symbol. These similar samples in the \ac{CP} and the data portion of the \ac{OFDM} symbol are spaced by $N$ samples apart. Using a sliding window correlator, this information can be used to detect the start of the \ac{OFDM} symbol. However, after applying the suppressing signal to the OFDM signal, the \ac{CP} samples are no longer a cyclic extension of the \ac{OFDM} symbol. As such, the \ac{CP} may no longer be utilized for synchronization purposes. To overcome this issue the suppressing signal can be designed so that it leaves part of the \ac{CP} and the corresponding samples in the data duration of the \ac{OFDM} symbol unaffected. Accordingly, part of the \ac{CP} samples are used by the suppressing signal for \ac{OOB} and \ac{PAPR} reduction while the rest are used for synchronization. This partial \ac{CP} usage is only during the synchronization phase, once synchronization is established the full \ac{CP} length can be utilized by the suppressing signal. 

Let $R$ denote the number of \ac{CP} samples used for synchronization located at the start of the \ac{OFDM} symbol. As mentioned above, the $R$ \ac{CP} samples as well as the corresponding $R$ data samples are not distorted in any way by the suppressing signal. As such, the transmitted signal during the synchronization period will be different than the one in \eqref{Eq:tx_signal}. We introduce the matrix $\mathbf{W}$ to preserve the \ac{CP} samples and their corresponding data samples as follows
\begin{align}
\mathbf{t}_{\rm s}=\mathbf{x}+\mathbf{c}_{\rm s}=\mathbf{AF}^{\rm H}\mathbf{Md+WP}\mathbf{s}_{\rm s}~,
\label{Eq:txsignal_sync}
\end{align}  
where $\mathbf{W} \in \mathbbm{R}^{(N+L)\times (N+L-2R)}$ and is constructed by selecting the $N+L-2R$ columns of $\mathbf{I}_{(N+L)}$ corresponding to the samples being protected from any distortion caused by the suppressing signal. Similar to \eqref{Eq:span_BH}, the alignment matrix $\mathbf{P}$ is designed such that $\text{span}\mathbf{(P)=\ker{(BHW)}}$. The only difference now is that $\text{rank}(\mathbf{(BHW)}) = N$, and accordingly $\dim{(\ker{\mathbf{(BHW)}})} = (N+L-2R)-\text{rank}(\mathbf{(BHW)}) = L-2R$. Therefore, $R< \frac{L}{2}$ for $\ker{\mathbf{(BHW)}}$ to exist. This practically means that the partial \ac{CP} samples cannot be larger than half of the full \ac{CP} length. Furthermore, compared to using the full $L$ \ac{CP} samples, the degrees of freedom utilized by the vector $\mathbf{s}_{\rm s}$ to suppress the spectrum and \ac{PAPR} of the transmitted signal in \eqref{Eq:txsignal_sync} are reduced to $L-2R$ during the synchronization phase. As such, this results in some degradation in the \ac{PAPR} and \ac{OOB} reduction performance. However, this performance loss is only during the synchronization phase and once synchronization has been established, performance will fall back to that of the full CP.
  
\section{Numerical Results}
In this section, we evaluate the \ac{OOB} reduction as well as the \ac{PAPR} performance of the proposed method with computer simulations. For simulation tractability, we consider an \ac{OFDM} system with $N=64$ subcarriers and a \ac{CP} length of $L=16$ samples. Additionally, we assume that the \ac{OFDM} transmitter detects an adjacent user spanning $10$ subcarriers within its band of transmission. Thus, these subcarriers are disabled by the \ac{OFDM} system, while the remaining subcarriers are utilized for transmission. The transmission is carried through a multipath Rayleigh fading channel with $L+1$ taps and a uniform \ac{PDP}. To illustrate the \ac{OOB} power leakage reduction performance of the proposed method, $10^4$ 4QAM symbols are generated randomly and the Welch's averaged periodogram method is then used to estimate the power spectrum. We evaluate the the PAPR reduction performance using the \ac{CCDF}. Furthermore, in all simulations, we constrain the power of the suppressing signal to be a fraction of the power of the plain \ac{OFDM} signal, i.e., $\epsilon=\alpha \mathbf{\|x\|}_2^2$, where $\alpha$ is a parameter that controls the power allocated to the suppressing signal. The maximum power percentage consumed by the suppressing signal is $\frac{\alpha}{1+\alpha}$ of the total available power budget. In all simulations, we assume the total power budget is shared between the \ac{OFDM} signal and the suppressing signal. 
\begin{figure}
\centering
\includegraphics[width=0.48\textwidth]{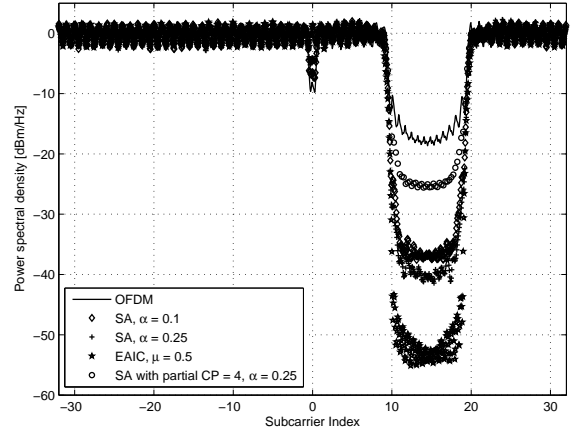}
\caption{Power spectral density for 4-QAM, $\lambda = 0$.}
\label{Fig:spectrum}
%\vspace{-5mm}
\end{figure}
\subsection{\ac{PAPR} and \ac{OOB} power leakage reduction performance}
First, we evaluate the \ac{OOB} power leakage reduction of the proposed method based on \eqref{Eq:optimization}, without considering the PAPR reduction, i.e., $\lambda = 0$. As shown in \figurename~\ref{Fig:spectrum}, the proposed method achieves remarkable levels of \ac{OOB} power leakage reduction compared to plain \ac{OFDM}. We also note that the amount of \ac{OOB} power leakage reduction increases as $\alpha$ increases, i.e., as more power is allocated to the suppressing signal. For example, for $\alpha = 0.1$ or approximately $\%9$ of the total power budget, the suppressing signal reduces the \ac{OOB} leakage by roughly $18$ dB, while approximately $22$ dB reduction is obtained when $\alpha=0.25$ or $\%20$ of the total power budget. We also note that the performance depends on the number of \ac{CP} samples used by the suppressing signal as demonstrated by the case of partial \ac{CP} usage in \figurename~\ref{Fig:spectrum}, where approximately $7$ dB reduction compared to \ac{OFDM} is achieved using only $4$ \ac{CP} samples. We note here that partial CP is used only during the synchronization phase. By examining \figurename~\ref{Fig:spectrum}, we observe a slight overshoot in the spectrum close to the band edges, especially as $\alpha$ grows. This can be attributed to the fact that the suppressing signal puts more power on the subcarriers close to the edges because of their high contribution to the \ac{OOB} power leakage. 
In \figurename~\ref{Fig:spectrum}, we also show the performance of the \ac{EAIC} scheme in \cite{5393050} evaluated under the same spectral efficiency as our proposed approach. Although the \ac{EAIC} scheme achieves better \ac{OOB} reduction compared to our proposed method, it does so by introducing distortion on the data subcarriers which leads to degradation in the \ac{BER} performance as will be shown below.

\begin{figure}
\centering
\includegraphics[width=0.48\textwidth]{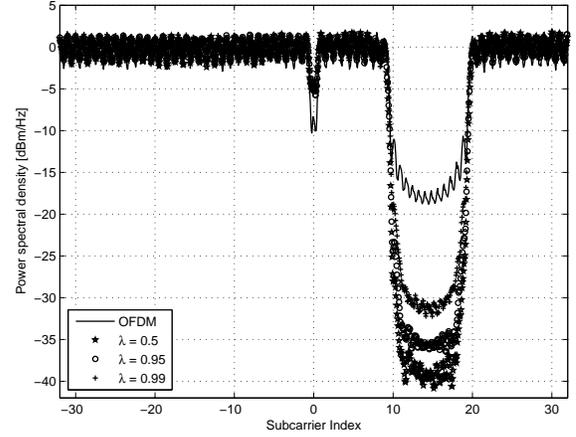}
\caption{Power spectral density for 4-QAM for $\alpha = 0.25$.}
\label{Fig:oob_lambda}
%\vspace{-5mm}
\end{figure}

\figurename~\ref{Fig:papr_oob_tradeoff} shows the trade-off between the \ac{PAPR} reduction and \ac{OOB} reduction performance for the joint optimization problem in \eqref{Eq:joint}. The trade-off is visualized by showing the average reduction in both \ac{OOB} leakage and \ac{PAPR} as a function of the adaptation parameter $\lambda$ when $\alpha$ is set to $0.25$. We note that when $\lambda=0$, the optimization problem \eqref{Eq:joint} is equivalent to \eqref{Eq:optimization}, where only the \ac{OOB} interference is minimized. The average reduction in \ac{OOB} interference in this case is approximately $22$ dB, which agrees with the results in \figurename~\ref{Fig:spectrum} when $\alpha=0.25$. Increasing $\lambda$ beyond zero, reduces the gain in terms of \ac{OOB} leakage reduction while gradually improving the \ac{PAPR} reduction performance. As shown in \figurename~\ref{Fig:papr_oob_tradeoff}, a maximum average \ac{PAPR} reduction of more than $3$ dB is obtained when $\lambda=1$. However, in this case, and as expected, there is no gain in the \ac{OOB} interference reduction. In fact, the \ac{OOB} power leakage increases due to the fact that the suppressing signal places some power in the adjacent band. The same is true when $\lambda = 0$, where a pure \ac{OOB} leakage reduction leads to a slight increase in the \ac{PAPR} as shown in \figurename~\ref{Fig:papr_oob_tradeoff}. 
\begin{figure}
\centering
\includegraphics[width=0.48\textwidth]{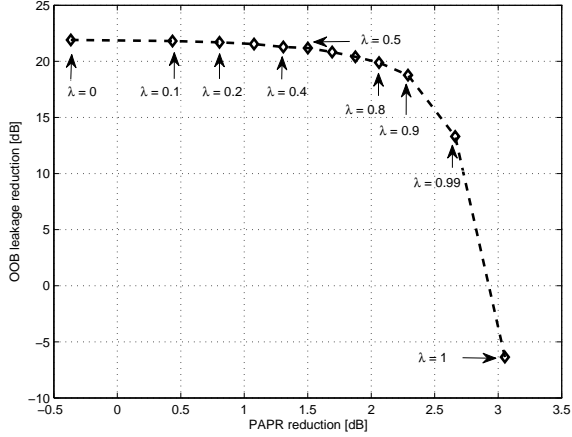}
\caption{Performance trade-off between \ac{OOB} reduction and \ac{PAPR} reduction when $\alpha = 0.25$.}
\label{Fig:papr_oob_tradeoff}
%\vspace{-5mm}
\end{figure}
The \ac{OOB} power leakage reduction for different values of $\lambda$ is shown in \figurename~\ref{Fig:oob_lambda}. Here, the power of the suppressing signal is fixed at $20\%$ of the total power budget, i.e., $\alpha=0.25$. These results in \figurename~\ref{Fig:oob_lambda} expand over the mean \ac{OOB} reduction results in \figurename~\ref{Fig:papr_oob_tradeoff} by showing the actual power spectral density of the transmitted signal. As seen from \figurename~\ref{Fig:oob_lambda}, the \ac{OOB} leakage is significantly reduced as $\lambda$ decreases, which is rather expected as more emphasis is put on the \ac{OOB} leakage reduction relative to the \ac{PAPR} reduction.
\begin{figure}
\centering 
\includegraphics[width=0.48\textwidth]{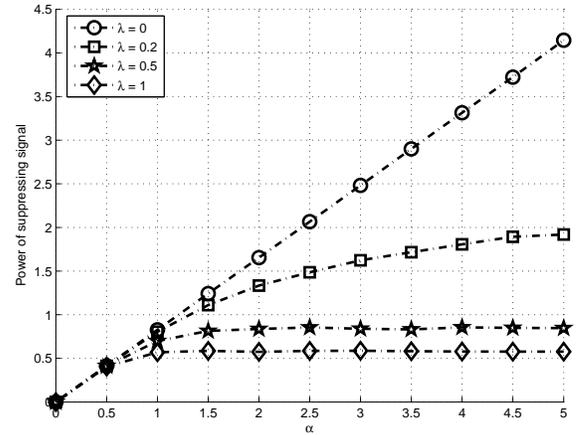}
\caption{Power of suppressing signal for different values of $\lambda$.}
\label{Fig:alpha_vs_Ps_power}
%\vspace{-5mm}
\end{figure}
In order to understand the behavior of the joint optimization problem in \eqref{Eq:joint} with regard to the actual power allocated to the suppressing signal, we plot the average power of the suppressing signal against $\alpha$ for different values of the adaptation parameter $\lambda$, as shown in \figurename~\ref{Fig:alpha_vs_Ps_power}. The results in \figurename~\ref{Fig:alpha_vs_Ps_power} indicate that when \ac{PAPR} is not considered, i.e., $\lambda=0$, the actual power used by the suppressing signal changes linearly with $\alpha$. In other words, all the power allocated to the suppressing signal will be completely utilized to reduce the spectral sidelobes. However, as the \ac{PAPR} reduction is slowly factored into the problem, the utilization of the allocated power decreases. Specifically, we observe that as the adaptation parameter $\lambda$ increases gradually, the suppressing signal uses less power to jointly reduce both \ac{PAPR} and spectral leakage. For the extreme case of $\lambda=1$, i.e., when it is a pure \ac{PAPR} reduction problem, the power of the suppressing signal completely saturates regardless of how much power is allocated through the parameter $\alpha$. 

\begin{figure}
\centering
\includegraphics[width=0.48\textwidth]{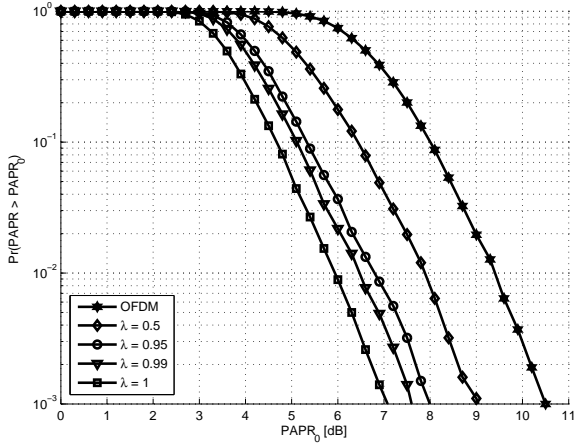}
\caption{\ac{PAPR} performance for 4-QAM when $\alpha = 0.25$.}
\label{Fig:papr_lambda}
%\vspace{-5mm}
\end{figure}

\begin{figure}
\centering
\includegraphics[width=0.48\textwidth]{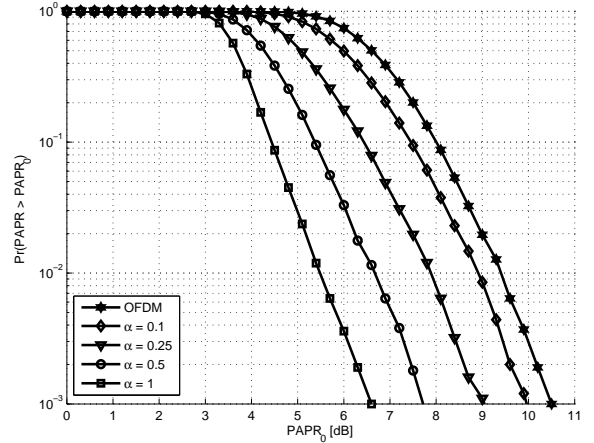}
\caption{\ac{PAPR} performance for 4-QAM when $\lambda = 0.5$.}
\label{Fig:papr_alpha}
%\vspace{-5mm}
\end{figure}

We now turn to characterizing the performance of the proposed method with regard to \ac{PAPR} reduction. In order to do that, we consider the actual instantaneous power distribution of the transmitted signal and plot its \ac{CCDF} as shown in \figurename~\ref{Fig:papr_lambda} and \figurename~\ref{Fig:papr_alpha}. In \figurename~\ref{Fig:papr_lambda}, we show the \ac{PAPR} performance for different values of $\lambda$ and a fixed $\alpha = 0.25$. We start by noting that remarkable reduction in the \ac{PAPR} is obtained as shown in \figurename~\ref{Fig:papr_lambda}. In particular, this reduction increases as the adaptation parameter $\lambda$ grows, i.e., the \ac{PAPR} reduction is emphasized compared to the power leakage reduction. For example, in the extreme case of $\lambda=1$, the \ac{PAPR} of the transmitted signal is around $7$ dB at a probability of $10^{-3}$; a reduction of approximately $3.5$ dB from that of the plain \ac{OFDM} signal. However, there is no reduction in the \ac{OOB} interference when $\lambda=1$. Nonetheless, decent improvements in the \ac{PAPR} performance can still be obtained even for small values of $\lambda$ while simultaneously allowing large reductions in the \ac{OOB} interference. For example, when $\lambda~\text{is set to}~0.5$, the \ac{PAPR} of the transmitted signal is around $9$ dB compared to $10.5$ dB at a probability of $10^{-3}$ for plain \ac{OFDM}. At the same value of $\lambda$, the \ac{OOB} power of the transmitted signal is around $-39$ dB compared to $-18$ dB for plain \ac{OFDM}; a $21$ dB reduction as shown in \figurename~\ref{Fig:oob_lambda}.
The variation of the \ac{PAPR} performance with the suppressing signal power is shown in \figurename~\ref{Fig:papr_alpha}. In the extreme case of having a suppressing signal consuming $\%50$ as the total power budget, i.e., when $\alpha=1$, the \ac{PAPR} is reduced by $4$ dB at a probability of $10^{-3}$. Alternatively, for $\alpha=0.25$, the \ac{PAPR} is reduced by approximately $1.5$ dB, showing that a slight increase in the power allocated to the suppressing signal can still lead to good \ac{PAPR} reduction.
\subsection{Bit error performance}
\begin{figure}
\centering
\includegraphics[width=0.48\textwidth]{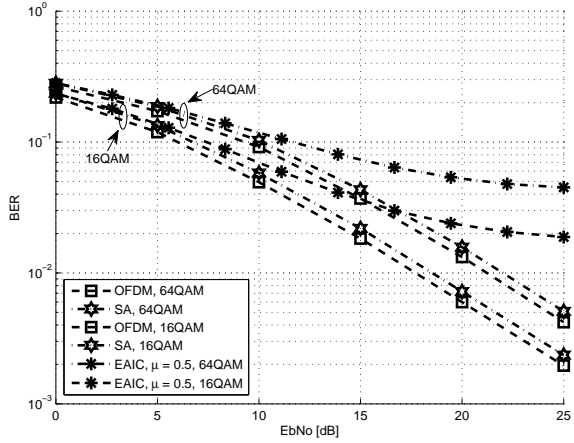}
\caption{\ac{BER} performance under Rayleigh multipath fading channels, $\lambda=0.5,~\alpha=0.25$.}
\label{Fig:ber_performance_SA_EAIC}
%\vspace{-5mm}
\end{figure}

The \ac{BER} performance of the proposed method as well as the \ac{EAIC} scheme \cite{5393050} for $16$QAM and $64$QAM is shown in \figurename~\ref{Fig:ber_performance_SA_EAIC}. The performance is evaluated in a Rayleigh multipath fading channel. The \ac{EAIC} clearly has an error floor due the distortion it introduces to the data subcarriers. On the contrary, our proposed appraoch offers a distortion-free transmission without any changes to the receiver structure. It's worth noting here that the small offset in performance between the proposed scheme and plain \ac{OFDM} is due to the fact that the total power budget is shared between the suppressing signal and the \ac{OFDM} signal.

\begin{figure}
\centering
\includegraphics[width=0.48\textwidth]{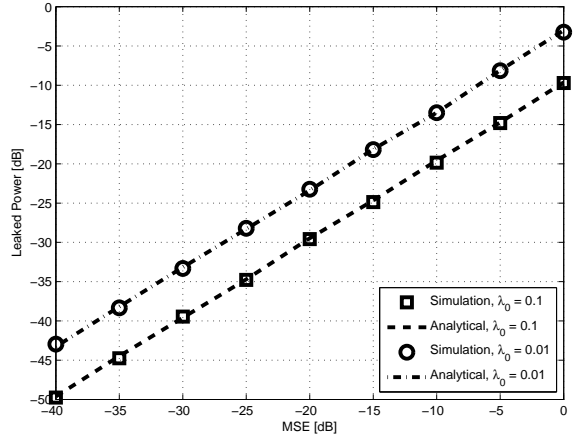}
\caption{Leaked power in \eqref{Eq:leaked_closed_form} under imperfect \ac{CSI}, $\lambda=0$.}
\label{Fig:leaked_power_analytical}
%\vspace{-5mm}
\end{figure}

\begin{figure}
\centering
\includegraphics[width=0.48\textwidth]{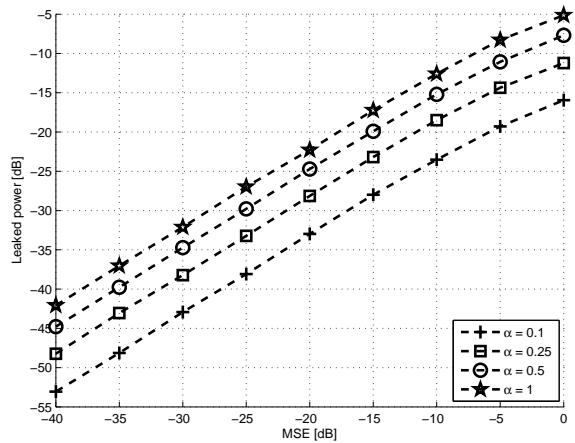}
\caption{Leaked power of the suppressing signal under imperfect \ac{CSI}, $\lambda=0.5$.}
\label{Fig:mse_vs_leaked_power}
%\vspace{-5mm}
\end{figure}

The \ac{BER} results in \figurename~\ref{Fig:ber_performance_SA_EAIC} are obtained with the assumption that there is no channel estimation errors. However, and as mentioned before if the correct channel is not perfectly known at the transmitter, the suppressing signal will leak into the OFDM signal. In \figurename~\ref{Fig:leaked_power_analytical}, we show the analytical leaked power expression in \eqref{Eq:leaked_closed_form} as well the simulated leaked power plotted against the \ac{MSE} of different channel errors when we consider the \ac{OOB} interference reduction only, i.e., $\lambda = 0$. It is clear that the leaked power values obtained from the closed-form expression in \eqref{Eq:leaked_closed_form} match those obtained from the simulation. The leaked power when we jointly consider the \ac{OOB} and PAPR reduction is shown in \figurename~\ref{Fig:mse_vs_leaked_power}, where the results are obtained through simulation since there is no closed-form expression for the leaked power in this case.
 
\begin{figure}
\centering
\includegraphics[width=0.48\textwidth]{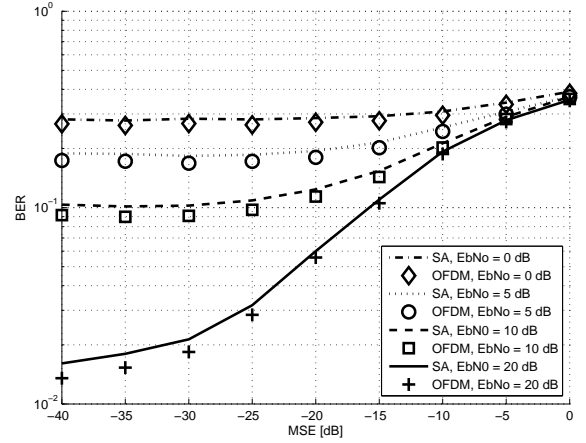}
\caption{\ac{BER} performance for $64$-QAM under imperfect \ac{CSI}, $\lambda=0.5,~\alpha=0.25$.}
\label{Fig:mse_vs_ber}
%\vspace{-5mm}
\end{figure}

To assess the impact of the power leakage due to the channel estimation errors on the \ac{BER} performance, we conducted Monte Carlo simulations of the proposed algorithm for different values of the \ac{SNR} as shown in \figurename~\ref{Fig:mse_vs_ber}. Our simulations are bench-marked against the error performance of plain \ac{OFDM} under the same noisy channel estimation. Other than a small offset as observed in \figurename~\ref{Fig:mse_vs_ber} which is due the total power budget being shared, the error performance of the suppressing alignment algorithm is identical to that of standard \ac{OFDM} under channel estimation errors. This can be explained by looking at \figurename~\ref{Fig:mse_vs_leaked_power}. There, the average power leakage of the suppressing signal into the data part of the received \ac{OFDM} symbol defined in \eqref{Eq:leaked_pow} 
is plotted against the \ac{MSE} of the channel for different values of $\alpha$. In \figurename~\ref{Fig:mse_vs_leaked_power}, the leaked power is at least $8$ dB less than the channel \ac{MSE} when $\alpha = 0.25$. Essentially, the noisy channel dominates the error performance. As such, no degradation in the  \ac{BER} is observed as shown in \figurename~\ref{Fig:mse_vs_ber} when $\alpha = 0.25$.
\section{Conclusion}
In this work, we have proposed an approach called {\em suppressing alignment} that generates a suppressing signal to jointly reduce the \ac{OOB} power leakage and \ac{PAPR} of \ac{OFDM}-based systems. The main advantage of the proposed method is that it does not reduce the spectral efficiency as it exploits the inherent redundancy in \ac{OFDM} provided by the \ac{CP}. We have also shown that the suppressing signal can be constructed in such a way that it does not create any interference to the information data carried in the \ac{OFDM} symbol. In particular, by utilizing the wireless channel, the suppressing signal is aligned with the \ac{CP} duration at the receiver, effectively creating an interference-free transmission with a \ac{BER} performance similar to legacy \ac{OFDM} without requiring any change in the receiver structure. The effectiveness of the proposed approach in obtaining remarkable reduction in both the \ac{OOB} power leakage and \ac{PAPR} is shown through computer simulations. We showed the performance trade-off between the \ac{OOB} power leakage reduction and \ac{PAPR} reduction where both can flexibly be controlled through an adaptation parameter. Furthermore, we investigated the impact of imperfect \ac{CSI} on the error performance of the proposed approach. Simulation results show no degradation in the \ac{BER} performance of the proposed approach compared to legacy \ac{OFDM} under the same noisy channel errors.

%\begin{IEEEbiography}{Anas Tom}
%xxx.
%\end{IEEEbiography}

%\begin{IEEEbiography}{Alphan Sahin}
%xxx.
%\end{IEEEbiography}

%\begin{IEEEbiography}{Huseyin Arslan}
%xxx.
%\end{IEEEbiography}

%\begin{IEEEbiographynophoto}{H\"{u}seyin Arslan}
%xxx.

%\newpage



%\vfill
\balance
\bibliographystyle{IEEEtran}
\bibliography{Tom_tcom-1038_Ref}

\end{document}